\documentclass{ifacconf}
\usepackage{graphicx}      % include this line if your document contains figures
\usepackage{natbib}  
\usepackage{amsmath,mathtools,amssymb}

\newenvironment{proof}[1][Proof]{\textbf{#1.} }{\ \rule{0.5em}{0.5em}}

\usepackage{amsfonts}
\usepackage{graphicx}
\usepackage{color}
\usepackage{float}
\usepackage{tabularx}
\usepackage{gensymb}
\usepackage{amssymb}
\usepackage{array}
\usepackage{amsmath,mathtools}
\usepackage{bm}
\usepackage{mathrsfs}

\usepackage{amssymb}
\newcommand{\sign}{\operatorname{sign}}
\usepackage{graphicx}
\usepackage{epsfig}
\usepackage{subfig}
\usepackage{float}

\newcommand{\Endofremark}{\@endgadget{\@Endofsymbol}}
\usepackage{multirow}
\usepackage{longtable}
\usepackage{booktabs}
\usepackage{tikz,pgfplots,pgfmath} % for drawings and plots, also loads package graphicx
\usetikzlibrary{math,positioning,intersections,shapes.misc,shapes.geometric,fit,arrows,arrows.meta,decorations.pathmorphing,decorations.pathreplacing,decorations.shapes,decorations.markings,patterns,calc,shadings}

\newtheorem{theorem}{Theorem}
\newtheorem{lemma}{Lemma}

\newtheorem{proposition}{Proposition}
\newtheorem{remark}{Remark}[section]

\newtheorem{assumption}{Assumption}
\newtheorem{definition}{Definition}
\newcommand{\R}{\mathbb{R}}
\begin{document}
\begin{frontmatter}
\title{ Robust safety design for strict-feedback nonlinear systems via observer-based linear time varying feedback}
\author[ImViA,COMSATS]{Imtiaz Ur Rehman}
\author[LIS]{Moussa Labbadi}
\author[ImViA]{Amine Abadi}
\author[ImViA]{Lew Lew Yan Voon}

\thanks[footnoteinfo]{We thank the French government for the Plan France Relance initiative which provided fundings via the European Union under contract ANR-21-PRRD-0047-01. We are also grateful to the Doctoral School and the French Ministry of Research for the PhD MENRT scholarship.}

\address[ImViA]{ImViA Laboratory EA 7535, Université Bourgogne Europe,  
720 avenue de l'Europe, 71200 Le Creusot, France  
(e-mail: Imtiaz-Ur.Rehman@ube.fr; Amine.Abadi@ube.fr; lew.lew-yan-voon@ube.fr)}

\address[COMSATS]{COMSATS University Islamabad, Islamabad, Pakistan  
(e-mail: imtiaz.rehman@comsats.edu.pk)}

\address[LIS]{Aix-Marseille University, LIS UMR CNRS 7020, 13013 Marseille, France  
(e-mail: moussa.labbadi@lis-lab.fr)}

\begin{abstract}
This paper develops a robust safety-critical control method for nonlinear strict-feedback systems with mismatched disturbances. Using a state transformation and a linear time-varying disturbance observer, the system is converted into a form that enables safe control design. The approach ensures forward invariance of the safety set and also applies to disturbance-free systems. Safety is proven for all cases, and a numerical example illustrates the results.
\end{abstract}
\begin{keyword}
Robust control, Safety-critical systems, Nonlinear systems, Forward invariance
\end{keyword}

% \keywords{}
\end{frontmatter}
\section{Introduction}
Safety-critical control has become essential for autonomous and robotic systems operating under uncertainty. Control Barrier Functions (CBFs), introduced for real-time safety filtering in \citep{ames2016control}, have been extended to robust and adaptive formulations for uncertain nonlinear systems \citep{xu2015robustness, buch2021robust, alan2022disturbance}. Recent developments include reduced-order safety design \citep{cohen2024safety}, safe backstepping \citep{taylor2022safe}, and applications to nonholonomic vehicles \citep{haraldsen2024safety} and robotic systems \citep{cortez2019control}.

Prescribed-time and finite-time safety frameworks have also emerged to ensure constraint satisfaction within a user-defined time horizon. These include prescribed-time safety for integrator and strict-feedback systems \citep{abel2022prescribed, abel2023prescribed}, as well as analyses of fundamental performance and robustness limitations \citep{aldana2023inherent}. Robust prescribed-time and fixed-time algorithms have been further explored in safety-critical contexts \citep{garg2021robust, nasab2025safe}.

Ensuring robust, fast, and computationally efficient safety-critical control under mismatched disturbances remains a significant challenge. This paper develops a robust safety-critical control framework for nonlinear strict-feedback systems. The proposed approach leverages a state transformation technique, originally introduced in \citep{praly1997generalized} and extended in \citep{abel2023prescribed} for free-disturbances, to convert strict-feedback dynamics into a chain-of-integrators structure suitable for safety analysis. Systems subject to both matched and mismatched disturbances are considered, and a disturbance observer is designed to estimate these uncertainties. A linear time-varying control law is then employed to synthesize a safety-critical controller for the transformed system. The method also applies to the nominal disturbance-free case. Rigorous analysis establishes forward invariance of the safety set under all considered scenarios, and a numerical example illustrates the effectiveness of the proposed strategy.

\textit{Notations}: This paper will utilize the notation listed below.  \( \mathbb{R}_+ = \{x \in \mathbb{R} : x \geq 0\} \) and \( \mathbb{R}^*_+ = \{x \in \mathbb{R} : x > 0\} \), where \( \mathbb{R} \) represents the set of real numbers. The absolute value in \( \mathbb{R} \) is denoted by \( | \cdot | \), and \( \| \cdot \| \) represents the Euclidean norm on \( \mathbb{R}^n \). For a (Lebesgue) measurable function \( d : \mathbb{R}_+ \to \mathbb{R}^m \), the norm \( \|d\|_\infty = \text{ess sup}_{t \ge 0} \|d(t)\| \) is defined. The set of functions \( d \) satisfying \( \|d\|_\infty < +\infty \) is denoted as \( L^\infty_m \). A function $\alpha$ : $\R^{+}$ $\rightarrow$ $\R^{+}$ is a class $\kappa$ provided $\alpha(0)$=0, strictly increasing and continuous; if $\alpha(\infty)=\infty$ and is of class $\kappa$ then function $\alpha$ is a class $\kappa^{e}_{\infty}$.
\vspace{-0.5cm}
\section{Background}
Consider the following system:
\begin{small}
 \begin{equation}\label{eq:disturbed_system}
    \dot{x} = f(x,t) + g(x,t)u + d(t), \quad x \in \mathbb{X} \subset  \mathbb{R}^n, \; u \in \mathbb{U} \subset \mathbb{R}^m,
\end{equation}   
\end{small}
where $t \in \mathbb{T} \triangleq [t_0, \infty)$ denotes the time with initialization $t_0\geq 0$, while $f : \mathbb{R}^n \times \mathbb{T} \to \mathbb{R}^n$ and $g : \mathbb{R}^n \times \mathbb{T} \to \mathbb{R}^{n \times m}$ are assumed locally Lipschitz in \(x\) and $d(t)$ denotes the external disturbance and assumed to be bounded as $\|d(t)\|_\infty:= \sup_{t \geq 0} \|d(t)\| \le \rho$, with $\rho>0$. The disturbance free definition of CBFs can be found in~\citep{ames2016control}.
% Evaluating $\dot{h}$ along \eqref{eq:disturbed_system} yields:
% \[
% L_{f}h(x) + L_{g}h(x)u + h_x(x)\,d
% \]
% The disturbance prevents us from looking for the condition
% $\dot{h}\geq - \Gamma(h)$, cf.~\eqref{eq:3}. This motivates the introduction of robust
% control barrier functions (RCBFs) as follows. 
\begin{definition}\label{def:2}~\citep{zhao2020adaptive}
A function $h:\mathbb{X}\to\mathbb R$ belongs to $\mathcal C^{n}$ is a robust CBF (RCBF) for the system (\ref{eq:disturbed_system}), on the safe set $\mathcal{S}$ 
 \begin{align}\label{eq:2}
\mathcal{S} &= \{x \in \mathbb{X} \subset  \mathbb{R}^n \,\,|\,\, h(x) \geq 0\},
\end{align}
such that $\forall x \in \mathcal{S}$ and $\forall t \geq 0$ :
\begin{equation}\label{eq:5} 
   \sup_{u \in U} \big[\underbrace{{L_{f}h(x)} + L_{g}h(x)u - \|h_x(x)\| \rho}_{\dot{h}(x,u,\rho)} \big]  \geq - \Gamma(h(x)),   
\end{equation}
where $\Gamma \in \kappa^{e}_{\infty}$, $L_f h(x) = \left( \frac{\partial h}{\partial x} \right) f(x)$, $L_g h(x) = \left( \frac{\partial h}{\partial x} \right) g(x)$ and $h_x(x) = \frac{\partial h(x)}{\partial x}$. Defining the point-wise control command set:
\begin{equation} \label{eq:rcbf_set}
    K_{\text{RCBF}}(x) \triangleq \left\{u \in \mathcal{U} \mid \dot{h}(x,u,\rho) \geq - \Gamma(h(x)) \right\}.
\end{equation}
% A set $\mathcal{S}$ is said to be forward invariant for~\eqref{eq:sys_gen_1} if for every $x(t_0) \in \mathcal{S}$, it holds that $x(t) \in \mathcal{S}$  $\forall t \geq t_0$, where $t_0\geq 0$ is the initialization time. Consequently, the system \eqref{eq:disturbed_system} is said to be safe.
\end{definition}
The aforementioned definition can encompass several types of uncertainty/disturbance owing to the generalized structure presented in~\citep{alan2025generalizing}.
% The generalized framework introduced in~\citep{alan2025generalizing} allows the above definition to encompass various forms of uncertainty/disturbance.
In this paper, we will used the following lemmas. 
\begin{lemma}\citep{levant2002universal}\label{eq:lemma_1} Examine the nonlinear differential equation with perturbation:
\begin{equation}\label{eq:lemma1_dyn}
\dot{Z} + \iota_1\,|Z|^{1/2}{\sign}\!\big(Z\big)
+ \iota_2 \int_0^t {\sign}\!\big(Z\big)\,d\tau \;=\; \Theta(t),
\end{equation}
where $Z$ represents the solution, $\iota_1,\iota_2>0$ are design gains, and $\Theta(t)$ is a bounded perturbation with $\|\dot\Theta(t)\|_\infty \le C$ for some known $C\ge0$. If the gains are chosen such that $\iota_1 \;\ge\; 1.5\sqrt{C}$ and $\iota_2 \;\ge\; 1.1\,C$, then $Z \rightarrow 0$ and $\dot{Z} \rightarrow 0$ in finite time. 
% Moreover, the convergence time $T_r$ is given by
% \begin{equation}
%  T_r \leq \frac{7.6\chi(0)}{\eta_2-C}   
% \end{equation}
% where $\chi(0)$ is the zero initial value of $\chi$.
\end{lemma}

\begin{lemma}\citep{levant1998robust}\label{eq:lemma_2}
 The first-order sliding-mode differentiator
\begin{align} \label{eq:smd}
% \dot\rho_0(t) &= \zeta_0= \rho_1(t) - \varepsilon_0\,|\rho_0(t)-f(t)|^{1/2}\operatorname{\sign}\!\big(\rho_0(t)-f(t)\big),\nonumber\\
% \dot\rho_1(t) &= -\varepsilon_1\,\operatorname{\sign}\!\big(\rho_1(t)-\rho_0(t)\big), \nonumber\\
\dot{\chi}_{0} &=\gamma_0= -\lambda_{1}\,|\chi_{0}-f(t)|^{\frac12}{\sign}(\chi_{0}-f(t))
+ \chi_{1}, \nonumber\\
\dot{\chi}_{1}
&= -\lambda_{2}\,{\sign}(\chi_{1}-\gamma_0)
\end{align}
where $f(t)$ is a known signal, $\lambda_{1},\lambda_{2} >0$ are tuning gains, and $\gamma_0$, ${\chi}_{0}$, ${\chi}_{1}$ are the states of \eqref{eq:smd}. If the initial deviations $\chi_{0}-f(t_0)$ and $\gamma_0-\dot f(t_0)$ are constrained, then $\zeta_0$ estimates $\dot f(t)$ with arbitrary precision.
 % converges to $\dot f(t)$ with an arbitrarily small residual (tunable by $\varepsilon_0,\varepsilon_1$). In practice, $\zeta(t)=\dot f(t)+\omega(t)$ where the error $\omega(t)$ can be made small and bounded.
   
\end{lemma}

\section{Problem statement}\label{sec:PF}
We next focus on perturbed vector strict-feedback systems, generalizing the affine dynamics in \eqref{eq:disturbed_system}, of the form
\begin{align}\label{eq:1}
\dot{\bf{x}}_i(t) &= {\bf{x}}_{i+1}(t) + \phi_i(\bar{\bf{x}}_{i}(t)) + d_i(t), \quad 1\leq i \leq n-1,\nonumber\\
\dot{\bf{x}}_n(t) &= G(\bar{\bf{x}}_{n}(t))u(t) + \phi_n(\bar{\bf{x}}_{n}(t)) + d_n(t),\nonumber\\
y(t) &= h(x_1(t)), \qquad t\ge 0
\end{align}
where ${\bf{x}}_i \in \mathbb{R}^{m}$ are vector valued states, and $\bar{\bf{x}}_{i}$ denotes the column vector as $\bar{\bf{x}}_{i}= [{\bf{x}}_1,...,{\bf{x}}_i]^\top \in\mathbb{R}^{mi}$ with ${\bf{x}}=\bar{\bf{x}}_{n}$ and $u\in \mathbb{R}^{m}$ is the
Lebesgue-integrable input. The functions  $G:\mathbb{R}^{mn} \rightarrow \mathbb{R}^{m \times m}$ and $\phi_i:\mathbb{R}^{mi} \rightarrow \mathbb{R}^{m}$ are defined as 
\begin{gather*}
G(\bar{\mathbf{x}}_{n}) := 
\begin{bmatrix}
g_{1,1}(\bar{\mathbf{x}}_{n}) & \cdots & g_{1,m}(\bar{\mathbf{x}}_{n})\\
\vdots & \ddots & \vdots\\
g_{m,1}(\bar{\mathbf{x}}_{n}) & \cdots & g_{m,m}(\bar{\mathbf{x}}_{n})
\end{bmatrix}, \;
\phi_i(\bar{\mathbf{x}}_{i}) := 
\begin{bmatrix}
\phi_{i,1}(\bar{\mathbf{x}}_{i})\\
\vdots\\
\phi_{i,m}(\bar{\mathbf{x}}_{i})
\end{bmatrix}.
\end{gather*}
 The disturbance vector  \( d(t) = [d_1(t) \ldots d_n(t)]^\top \in \mathbb{R}^n \) representing the mismatched and matched disturbances.
% \begin{equation}\label{eq:1}
%     \begin{aligned}
% \dot{x}_i(t) &= x_{i+1}(t) + \phi_i(\bar{x}_i(t)) + d_i(t), \quad 1\leq i \leq n-1,\\
% \dot{x}_n(t) &= u(t) + \phi_n(x(t)) + d_n(t),\\
% y(t) &= h(x_1(t)), \qquad t\ge 0
% \end{aligned}
% \end{equation}
% where $\bar{x}_i=[x_1,...,x_i]^T\in \mathbb{R}^{i}$, ${x}=[x_1,...,x_n]^T$, $y\in \mathbb{R}$, $u\in \mathbb{R}$ are the state, output and control input of the system, respectively. The disturbance vector is given by \( d(t) = [d_1(t) \ldots d_n(t)]^\top \in \mathbb{R}^n \) representing the mismatched and matched disturbances.

% \end{ass}
\begin{assumption}\label{eq:ass_1}
    The disturbances are bounded and
satisfies $\|d(t)\|\leq \rho$, $\rho>0$.
\end{assumption}
\begin{assumption}\label{eq:ass_2}
    The matrix $G(\bar{\bf{x}}_{n})$ is non-singular $\forall \bar{\bf{x}}_{n} \in \mathbb{R}^{mn}$. Moreover, for each $i = 1,. . ,n$, the functions $\phi_{i,k}(\bar{\bf{x}}_{i})$, $k = 1,. . ., m$ are $n-i$ times differentiable.
\end{assumption}
Despite the structural differences between~\eqref{eq:1} and~\eqref{eq:disturbed_system}, the system can be transformed into an input-affine chain of integrators (see the next paragraph) via standard backstepping coordinate transformations. As a result, the safety notions and the CBF/RCBF conditions introduced earlier remain valid in the transformed coordinates, where the control input enters linearly through a virtual control variable. Hence, the same safety-critical control framework can be systematically extended to this well-known class of perturbed strict-feedback systems.
% {\color{blue}
% Although the structure of~\eqref{eq:1} differs from~\eqref{eq:disturbed_system}, the system can be transformed into an input-affine chain of integrators (as shown in the subsequent paragraph) 
% through backstepping coordinate transformations. Therefore, the safety definitions, CBF and RCBF conditions introduced in the previous section remain valid in the transformed domain, where the control input enters linearly through a virtual control variable.
% Consequently, the same safety-critical control framework can be systematically extended to this class of perturbed strict-feedback systems.}

\subsection{Coordinate Transformation}
This study aims to design a safe control law for the system (\ref{eq:1}), in the presence of both matched and mismatched disturbances. The design process in our work begins with a transformation of system (\ref{eq:1}) as follows:
\begin{gather*}
\boldsymbol{\beta}_0 = 0,\;  
\boldsymbol{\beta}_i(\bar{\bf{x}}_i) = -{\phi}_i(\bar{\bf{x}}_i)
+ \sum_{k=1}^{i-1} 
\frac{\partial {\boldsymbol{\beta}}_{i-1}}{\partial {\bf{x}}_k}
\big({\bf{x}}_{k+1} + {\phi}_k(\bar{\bf{x}}_k)\big), \\
\quad \quad \quad \quad 1\leq i \leq n.  
\end{gather*}
The introduce new coordinates
\begin{align}\label{eq:1_d}
{\boldsymbol{\varphi}}_i &= {\bf{x}}_i - {\boldsymbol{\beta}}_{i-1}(\bar{\bf{x}}_{i-1}), \qquad 1\leq i \leq n.    
\end{align}
with the input transformation is
\begin{align}\label{eq:1_e}
{u} &= G^{-1}(\bar{\bf{x}}_n)\big({v} + {\boldsymbol{\beta}}_n(\bar{\bf{x}}_n)\big).    
\end{align}
differentiating (\ref{eq:1_d}) and substituting the system dynamics (\ref{eq:1}) along with the input transformation and reformulating its dynamics gives a perturbed vector chain of integrators:
\begin{align}\label{eq:1_f}
\dot{\boldsymbol{\varphi}}_i &= {\boldsymbol{\varphi}}_{i+1}(t) + {{w}}_i(t,x), \quad 1\le i \le n-1 \nonumber\\
\dot{\boldsymbol{\varphi}}_n &= \boldsymbol{v}(t) + {w}_n(t,x),\nonumber\\
y(t)&=h(\boldsymbol{\varphi}_1(t)), \quad \forall t \geq 0
\end{align}
where
\begin{align}\label{eq:1_g}
{w}_i(t,x)&= {d}_i(t)
- \sum_{k=1}^{i-1}
\frac{\partial{\boldsymbol{\beta}}_{i-1}}{\partial {x}_k}\,{d}_k(t),\nonumber\\
{w}_n(t,x)&= {d}_n(t)
- \sum_{k=1}^{n-1}
\frac{\partial{\boldsymbol{\beta}}_{n-1}}{\partial {x}_k}\,{d}_k(t)
\end{align}
\begin{remark}
Note that the effects of all mismatched and matched disturbances accumulate in the residual term ${w}(t,x)$. The mismatch disturbances $d_k$ for $(k<i)$ persist because the terms $- \sum\frac{\partial{\boldsymbol{\beta}}_{i-1}}{\partial {x}_k}\,{d}_k$ propagate them to higher levels. If only the matched disturbance is present, i.e., ${d}_n \neq 0$ and $d_k=0$ $\forall k<n$, then ${w}_i=0$ $\forall i<n$ and ${w}_n={d}_n$.
% Observe that the residual ${w}(t,x)$ aggregates the impacts of all mismatched and matched disturbances. The mismatched disturbances $d_k$ for $(k<i)$ persist, as they are fed into higher levels via the terms $- \sum\frac{\partial{\boldsymbol{\beta}}_{i-1}}{\partial {x}_k}\,{d}_k$.
\hfill$\blacksquare$
\end{remark}
The terms 
$w_i(t,x)$ capture how the original mismatched and matched perturbations propagate through the backstepping change of coordinates, as shown in the subsequent example. 
% We take an example to illustrate it further.
\textbf{Example:} Consider the following strict-feedback system:  
\begin{gather*}
\dot{x}_1 = x_1^2 + x_2 + d_1,\;
\dot{x}_2 = x_1^2 + x_2^2 + x_3 + d_2,\;
\dot{x}_3 = u + d_3, \\ 
y = x_1,\;
\phi_1(x_1) = x_1^2,\;
\phi_2(\bar{x}_2) = x_1^2 + x_2^2,\;
\phi_3 \equiv 0,\\
G(\bar{x}_3) = 1, 
\beta_0 = 0,\;
\beta_1(x_1) = -x_1^2,\;
\beta_2(\bar{x}_2) = -x_1^2 - x_2^2 \\ - 2{x_1} x_2
- 2{x_1}^3\;  
\varphi_1 = x_1 + w_1,\;
\varphi_2 = x_2 + x_1^2 + w_2,\\
\varphi_3 = x_3 + x_1^2 + x_2^2  + 2x_1 x_2 
+ 2x_1^3 + w_3,\; w_1 = d_1,\; 
w_2 = d_2 \\ + 2{x_1} d_1,\;
w_3 = d_3 + (2{x_1} + 2{x_2} + 6{x_1}^2)d_1 
+ (2{x_2} + 2{x_1})d_2.
\end{gather*}
\begin{assumption}\label{ass:DO_W_bounds}
Consider the transformed system \eqref{eq:1_f} with $w_i(t,x)$ defined in~\eqref{eq:1_g}. 
Assume that for each $i\in\{1,\dots,n\}$ there exist finite constants $\bar w_i,\bar\delta_i>0$ such that, $\forall t\ge 0$:
$\|w_i(t,x)\|\le \bar w_i$, 
$\|\dot w_i(t,x)\|\le \bar\delta_i$.
\end{assumption}

{\color{blue}
}
Since ${\boldsymbol{\varphi}}_1 = \bf{x}_1$, the original output in \eqref{eq:1} and the transformed output in \eqref{eq:1_f} are identical. As a result, we focus the safety design on the transformed dynamics \eqref{eq:1_f} by synthesizing the input $v$, and then recover the actual input 
$u$ for the original strict-feedback system using the inverse mapping in \eqref{eq:1_e}.
 As the safety design is carried out in the transformed dynamics \eqref{eq:1_f}, any Lebesgue-integrable nominal control input $u_{no}$ formulated for the original system \eqref{eq:1} must be mapped into the transformed coordinates. This is accomplished by specifying 
\begin{align}\label{eq:1_h}
   v_{no} &= G(\bar{\bf{x}})u_{no} - {\boldsymbol{\beta}}_n(\bar{\bf{x}}_n)
\end{align}
as in \eqref{eq:1_h}. Our safety filter is thereafter applied directly to the converted nominal input $v_{no}$.

Our previous work \citep{labbadi2025robust} addressed only single-input linear integrator dynamics with disturbances, constant system matrices, and a fixed input channel. It cannot accommodate the nonlinear, state-dependent input matrix, coupled subsystem dynamics, or the propagation of mismatched disturbances present in strict-feedback systems, neither it support multi-input or vector-valued states. Furthermore, imposing constant upper bounds on the disturbances introduces conservatism. Thus, a disturbance observer is introduced to cope with this issue.
Prescribed-time safety and persistent safety address distinct control objectives. Prescribed-time methods such as \citep{abel2022prescribed, abel2023prescribed} guarantee safety only up to a fixed terminal time $t_f$ and assume disturbance free dynamics, limiting their robustness and practical applicability, particularly because their time-varying gains diverge as $t \rightarrow t_f$. In contrast, the persistent safety framework developed in this work ensures safety for all $t \geq 0$ under mismatched and matched perturbations. By using bounded, strictly increasing gains, the controller prevents singularities while guaranteeing robust forward invariance. This enables continuous safe operation in long-duration, safety-critical applications where robustness to uncertainties/disturbances is important, such as aerial robotics, autonomous navigation, and industrial process control. 
\section{Main results}\label{sec:main}
\subsection{Vector chain of integrators without disturbances}\label{eq:integrator_wo_d}
In this subsection, we first consider the unperturbed system case and aim to propose a control architecture that ensures the persistent safety of a vector chain of integrators in the absence of disturbances. Conventional CBF approaches are generally insufficient for systems with high relative degree (RD). To address this, we introduce a new RD-one CBF and synthesize the corresponding controller via a Quadratic Program (QP), leveraging the backstepping technique presented in \citep{krstic2006nonovershooting}. Specifically, consider system (\ref{eq:1_f}) with $w_i \equiv 0 $ and a candidate CBF $h_1(\boldsymbol{\varphi})$ of relative degree $n > 1$, which will serve as the basis for the design algorithm, under the following assumption:
\begin{assumption}\label{eq:ass_4}
The function $h_1(\boldsymbol{\varphi}):\mathbb R^n\to\mathbb R$ belongs to $\mathcal C^{n}$ and satisfies
\begin{equation}
    \frac{\partial h_1(\boldsymbol{\varphi})}{\partial \boldsymbol{\varphi}} \neq 0, \quad \forall \boldsymbol{\varphi} \in \hat{\mathcal{S}},
\end{equation}
whereas, the safe set $\hat{\mathcal{S}}$ is defined as:
\begin{align}\label{eq:2_d}
\hat{\mathcal{S}} &= \{\boldsymbol{\varphi}\in \mathbb{R}^{n} \;| \; h_1(\boldsymbol{\varphi}) \geq 0\},
\end{align}
with boundary $\partial \hat{\mathcal{S}} = \{\boldsymbol{\varphi} \in \mathbb{R}^{n} \;| \; h_1(\boldsymbol{\varphi}) = 0\}$ and interior $\text{Int}(\hat{\mathcal{S}}) = \{\boldsymbol{\varphi} \in \mathbb{R}^{n} \;| \; h_1(\boldsymbol{\varphi}) > 0\}$.
\end{assumption}
Computing the time derivative of $h_1(\boldsymbol{\varphi})$ gives
\begin{small}
\begin{align}\label{eq:8}
    \dot{h}_1(\boldsymbol{\varphi}) &= L_f h_1(\boldsymbol{\varphi}), \nonumber\\
                 &= -\bar{\varrho}_1 \Upsilon(t) h_1(\boldsymbol{\varphi}) + \underbrace{\bar{\varrho}_1 \Upsilon(t) h_1(\boldsymbol{\varphi}) + L_f h_1(\boldsymbol{\varphi})}_{h_2(\boldsymbol{\varphi})},
\end{align}    
\end{small}
where $\bar{\varrho}_1 >0$, and $\Upsilon(t)=1+t$ is a
strictly increasing time varying function. 
\begin{remark}\label{rm:3}
In \eqref{eq:8}, any continuous and strictly increasing function 
$\Upsilon : \mathbb{R}^{+} \to \mathbb{R}^{+}$ satisfying $\Upsilon(0) > 0$ 
and possessing an unbounded integral can be employed.  
For instance, one may choose exponential, linear, or polynomial functions, for example: 
\(
\Upsilon(t) = ae^{\alpha t}, \; a,\alpha>0, \; 
\Upsilon(t) = 1+t, \;
\Upsilon(t) = (1+t)^{p}, \; p>0.
\)
\hfill$\blacksquare$
\end{remark}
\begin{remark}\label{rm:4}
The prescribed-time form $\Upsilon(t)=\frac{\Upsilon_0}{T-t}$, with $\Upsilon_0>0$, $T>0$, $t\in[0,T)$, can enforce finite-time safety but suffers from singularity at $t=T$, which prohibits its usage for persistent safety applications, and also lack robustness owing to noise/disturbance amplification. 
On the other hand, the functions in Remark~\ref{rm:3} circumvent singularity and are recommended in persistent safety-critical applications. 
\hfill$\blacksquare$
\end{remark}
Using condition \eqref{eq:5}, the inequality $\dot{h}_1\geq - \Gamma(h_1)$ holds for $h_1(\boldsymbol{\varphi})$ in the disturbance-free case, given that $h_2(\boldsymbol{\varphi}) \geq 0$. As a consequence, a new candidate CBF given by
\begin{align}\label{eq:9}
    {h}_2(\boldsymbol{\varphi}) &= \bar{\varrho}_1 \Upsilon(t) h_1(\boldsymbol{\varphi}) + L_f h_1(\boldsymbol{\varphi}).
\end{align}
However, the initial condition must lie within the safe set defined by $h_2(\boldsymbol{\varphi})$ $(i.e., h_2(\boldsymbol{\varphi}_0)>0)$. To ensure this, we impose the subsequent assumption:
\begin{assumption}\label{eq:ass_3}
  $ h_1(\boldsymbol{\varphi}_0)>0$.
\end{assumption}

In standard CBF-based safety frameworks, it is typically considered that the initial condition satisfies $h_{1}(\boldsymbol{\varphi}_{0}) \geq 0$. However, we do not deal with situations in which the initial state is right on the boundary of the safe set $\hat{\mathcal{S}}$ (i.e., $h_{1}(\boldsymbol{\varphi}_{0}) = 0$), as ensuring safety from such places can pose control feasibility issues. Thus, under Assumption~\ref{eq:ass_3}, we select
\begin{equation}\label{eq:10}
    \bar{\varrho}_1 > \max\!\left\{ 0,\; -\,\frac{L_f h_1(\boldsymbol{\varphi}_0)}{\Upsilon(t_0)\,h_1(\boldsymbol{\varphi}_0)} \right\},
\end{equation}
this ensures $h_2(\boldsymbol{\varphi}_0)>0$ in accordance with (\ref{eq:9}). The scaling of the time-varying function $\Upsilon(t)$ at each stage, with its power proportional to the step index $i$, is a crucial design aspect. As a result, feedback improves over time, which speeds up convergence to the interior of the safe set and inhibits the propagation of disturbances along the integrator chain. Inspired by~\citep{abel2023prescribed,labbadi2024hyperexponential}, this scheme incorporates a significant time-varying feedback gain that offers strong safety for higher-order systems. The application of the backstepping transformation to system (\ref{eq:1_f}) with $w_i \equiv 0 $ is carried out as follows:
% An important design consideration is the scaling of the time-varying function $\Upsilon(t)$ at each transformation stage. Reinforcing the impact of $\Upsilon(t)$ at each stage is an effective approach for ensuring robustness and elevating performance for higher-order systems. In addition to expediting convergence to the inside of the safety set for the nominal system, this method provides a more powerful, time-varying feedback mechanism that is crucial for controlling the propagation of perturbations along the integrator chain. To accomplish this, $\Upsilon(t)$ is raised to a power proportional to the step index $i$. Inspired by~\citep{abel2023prescribed} and~\citep{labbadi2024hyperexponential}, this methodology offers robust safety guarantees for systems of any relative degree through the inclusion of a substantial time-varying feedback gain.
% Employing the backstepping transformation to the system (\ref{eq:1_f}) with $w_i \equiv 0 $ proceeds as follows:
\begin{align}\label{eq:11_a}
    h_1(\boldsymbol{\varphi}) &= h_1(\boldsymbol{\varphi}), \nonumber\\
    h_i(\boldsymbol{\varphi}) &= \bar{\varrho}_{i-1} \Upsilon(t)^{\vartheta(i-1)} h_{i-1}(\boldsymbol{\varphi}) + L_f h_{i-1}(\boldsymbol{\varphi}),
\end{align}
for $i=\{2, \ldots, n\}$, where the tuning factor $\vartheta \geq 1$, controls the intensity of the variable gain adjustment. Additionally, $\bar{\varrho}_{i-1}$ are selected to ensure that $h_i(\boldsymbol{\varphi}, t_0)$ is initially positive: 
\begin{equation}\label{eq:10_a}
   \bar{\varrho}_{i-1} > \max\!\left\{ 0,\; -\,\frac{L_f h_{i-1}(\boldsymbol{\varphi}_0)}{\Upsilon(t_0)^{\vartheta(i-1)}\,h_{i-1}(\boldsymbol{\varphi}_0)} \right\}.
\end{equation}
It is possible to explicitly specify the appropriate safety control law in (\ref{eq:13_a}) after constructing the CBF \(h_n(\boldsymbol{\varphi})\) with RD one. This law overrides the nominal control $\boldsymbol{v}_{\mathrm{no}}$ if the trajectories go close to the boundary of \(\hat{\mathcal{S}}\). Consequently,  the safety requirement is enforced by the final control $\boldsymbol{v}$, which ensures the forward invariance of $\hat{\mathcal{S}}$ and prevents violations:
\begin{align}\label{eq:13_a}  
\boldsymbol{v} &= \arg\min_{\boldsymbol{v}} \; \|\boldsymbol{v} - \boldsymbol{v}_{\mathrm{no}}\|^2 \nonumber \\ 
\text{s.t. } & {\dot{h}_n(\boldsymbol{\varphi},\boldsymbol{v})} \geq -\bar{\varrho}_{n} \, \Upsilon(t)^{\vartheta n}\, h_n(\boldsymbol{\varphi}),
\end{align}
where $\bar{\varrho}_{n}>0$. By employing the Karush-Kuhn-Tucker (KKT) optimality conditions, the closed-form solution to this QP is obtained, yielding the following expression~\citep{boyd2004convex}:
\begin{align}\label{eq:18_b} 
\boldsymbol{v} = \begin{cases} 
\boldsymbol{v}_{no}, & \zeta(\boldsymbol{\varphi}, \boldsymbol{v}_{no}) \geq 0, \\ 
\boldsymbol{v}_{no} - (L_g h_n(\boldsymbol{\varphi}))^\top \frac{\zeta(\boldsymbol{\varphi}, \boldsymbol{v}_{no})}{\|L_g h_n(\boldsymbol{\varphi})\|^2}, & \text{otherwise},
\end{cases}
\end{align}
wherein
\begin{align}\label{eq:19_a} 
\zeta(\boldsymbol{\varphi}, \boldsymbol{v}_{no}) &:= {\dot{h}_n(\boldsymbol{\varphi},\boldsymbol{v}_{no})} +\bar{\varrho}_{n} \, \Upsilon(t)^{\vartheta n}\, h_n(\boldsymbol{\varphi}).
\end{align}
\begin{remark}
    Our approach offers persistent safety, which is necessary in applications like continuous collision avoidance, in contrast to~\citep{abel2023prescribed}, which addresses temporary safety tasks. Furthermore, to ensure limited, monotonic gain growth and circumvent singularities, we opt a strictly increasing but bounded function instead of a blow-up function with unbounded gains and potential singularities. 
      \hfill$\blacksquare$
\end{remark}
\begin{proposition}\label{pro:2}
Consider system (\ref{eq:1_f}) with $w_i \equiv 0 $ that satisfies Assumptions~\ref{eq:ass_4} and~\ref{eq:ass_3}. Then the control law (\ref{eq:18_b}), synthesized employing the backstepping transformation (\ref{eq:11_a}) with the initial gain (\ref{eq:10_a}) and $\bar{\varrho}_n > 0$, ensures that $h_1(\boldsymbol{\varphi}(t))\geq 0$ $\forall t \in [t_0, \infty)$.
\end{proposition}

\subsection{Disturbance observer}\label{eq:DO}
In the previous subsection, the QP formulation ensures safety in the disturbance-free case. However, practical systems often encounter external disturbances. Considering a constant
bound for the perturbed system \eqref{eq:1_f} to
represent the time-varying disturbances in CBF constraints may lead to undesired conservativeness and a performance reduction in closed-loop. To cope with this, inspired by~\citep{chen2015robust}, we employed a disturbance observer to estimate the unmatched and matched external disturbances in~\eqref{eq:1_g}. The disturbance observer is designed as follows. Defining the error and its derivative as:
\begin{align}\label{eq:e1_dyn}
e_i &= \boldsymbol{\varphi}_i - \boldsymbol{\varphi}_d, \nonumber\\
\dot{e}_i &= \dot{\boldsymbol{\varphi}}_i - \dot{\boldsymbol{\varphi}_d} = \boldsymbol{\varphi}_{i+1} + w_i(t,x) - \dot {\boldsymbol{\varphi}}_d,
\end{align}
where $\boldsymbol{\varphi}_d$ is the desired value for $\boldsymbol{\varphi}_i$ in \eqref{eq:1_f}, we simply set $\boldsymbol{\varphi}_d = 0$ for safety. Note that to keep the observer design uniform for all $i=1,\dots,n$ in \eqref{eq:e1_dyn}, we adopt the convention $\boldsymbol{\varphi}_{n+1} = \boldsymbol{v}$. Then, to estimate the external disturbance $w_i(t,x)$, an auxiliary variable is defined as
\begin{align}\label{eq:s1_dyn}
\sigma_{i0} = r_i - e_i,\; 
\dot r_i = \boldsymbol{\varphi}_{i+1} + \hat w_i(t,x),   
\end{align}
whereas $\hat w_i$ is the estimate of $w_i$. Taking into account \eqref{eq:e1_dyn} and \eqref{eq:s1_dyn}, we obtain
\begin{align}\label{eq:sigma10_dot}
\dot{\sigma}_{i0} &= \hat w_i(t,x) - w_i(t,x)= \tilde w_i.     
\end{align}
Thus, the derivative of $\sigma_{i0}$ is exactly the disturbance estimation error. Since $\dot{\sigma}_{i0}$ is not measured directly, we approximate it using a first-order sliding mode differentiator according to lemma~\ref{eq:lemma_2}:
\begin{align}\label{eq:SMDO_1}
\dot{\chi}_{i0} &=\zeta_i= -\lambda_{i0}\,|\chi_{i0}-\sigma_{i0}|^{\frac12}{\sign}(\chi_{i0}-\sigma_{i0})
+ \chi_{i1}, \nonumber\\
\dot{\chi}_{i1}
&= -\lambda_{i1}\,{\sign}(\chi_{i1}-\zeta_i),
\end{align}
where $\lambda_{i0},\lambda_{i1}>0$ are the design gains, and $\dot{\chi}_{i0}$, $\dot{\chi}_{i1}$ and $\zeta_i$ are the states of \eqref{eq:SMDO_1}. As per \eqref{eq:SMDO_1} and lemma~\ref{eq:lemma_2}, we have
\begin{equation}\label{eq:fosmd}
 \dot{\sigma}_{i0} = \zeta_i + \Omega_i,
\end{equation}
where $\Omega_i$ is the differentiator residual. Defining $\sigma_{i1} = \sigma_{i0} + \zeta_i$, the disturbance observer is proposed as 
\begin{small}
 \begin{equation}\label{eq:fosmd_1}
 \dot{\hat w}_i = -\,\zeta_i - k_{i1}\,|\sigma_{i1}|^{\frac12}{\sign}(\sigma_{i1}) - k_{i2}\!\int_0^t {\sign}\big(\sigma_{i1}\big)\,d\tau,
\end{equation}   
\end{small}
By differentiating \eqref{eq:sigma10_dot}, we get $\ddot{\sigma}_{i0}=\dot{\hat{w}}_i - \dot{w}_i$. employing \eqref{eq:fosmd}, it follows that 
\begin{align}\label{eq:fosmd_2}
\dot{\sigma}_{i1} &= \dot{\sigma}_{i0} + \ddot{\sigma}_{i0}-\dot{\Omega}_i=\dot{\sigma}_{i0} + \dot{\hat{w}}_i - \dot{w}_i-\dot{\Omega}_i.
\end{align}
Invoking \eqref{eq:fosmd} and substituting
\eqref{eq:fosmd_1} into \eqref{eq:fosmd_2} we obtain
 \begin{align}\label{eq:fosmd_3}
 \dot{\sigma}_{i1}
&\;+\; k_{i1}\,|\sigma_{i1}|^{\frac12}{\sign}(\sigma_{i1})
\;+\; k_{i2}\!\int_0^t {\sign}(\sigma_{i1})\,d\tau \nonumber\\
&= \;\Omega_i - \dot{w}_i - \dot{\Omega}_i
= \bar{D}_i.
\end{align} 
\begin{proposition}\label{prop:DO_vector}
Consider the system in \eqref{eq:1_f}
with disturbances $w_i(t,x)$, $i=1,\dots,n$ in \eqref{eq:1_g}, satisfying Assumption~\ref{ass:DO_W_bounds}.
The disturbance observer given in \eqref{eq:fosmd_1} for $i\in\{1,\dots,n\}$ with $\bar D_i$
% \begin{equation}\label{eq:Di_bar_def}
% \bar D_i\;=\;\Omega_i-\dot w_i-\dot\Omega_i,
% \end{equation}
as given in \eqref{eq:fosmd_3}, where $\dot\Omega_i$ and $\dot w_i$ are the derivatives of the residual and the disturbance, respectively. Assume that there exists a known constant $\bar\varsigma_i>0$ such that
\begin{equation}\label{eq:Di_bar_bound}
\|\dot{\bar D}_i\|\le \bar\varsigma_i.
\end{equation}
If the observer gains satisfy, 
\begin{equation}\label{eq:k_gains_prop}
k_{i1} \ge 1.5\sqrt{\bar\varsigma_i},\qquad
k_{i2}\ge 1.1\bar\varsigma_i, \qquad \forall i=1,\dots,n,
\end{equation}
then the disturbance estimation errors $\tilde w_i=\hat w_i-w_i$ converge to zero
in finite time. 
% Whereas
% \begin{equation}\label{eq:wtilde_vec_def}
% \tilde w(t)\;:=\;\big[\,\tilde w_1^\top\;\tilde w_2^\top\;\cdots\;\tilde w_n^\top\,\big]^\top.
% \end{equation}
% there exists a finite $T>0$ such that
% \begin{equation}\label{eq:wtilde_vec_conv}
% \tilde w(t)=0,\qquad \forall t\ge t_0+T.
% \end{equation}
\end{proposition}

% auxiliary variable and sliding signal
% \begin{align}
% s_i^{\mathrm{DO}}(t) &:= z_i^{\mathrm{DO}}(t)-\varphi_i(t), \label{DO:si_def} \\
% \dot z_i^{\mathrm{DO}}(t) &=
% \begin{cases}
% \varphi_{i+1}(t) + \hat{\delta}_i^{\mathrm{DO}}(t), & i=1,\dots,n-1,\\[4pt]
% v(t) + \hat{\delta}_n^{\mathrm{DO}}(t), & i=n.
% \end{cases}
% \label{DO:zi_dyn}
% \end{align}
% % Explanation: differentiating s_i^{DO} gives the estimation error dynamics
% % \dot s_i^{DO} = \hat{\delta}_i^{DO} - w_i(t,x), so s_i^{DO} directly encodes the error.
% \begin{align}
% \dot r_{i0}(t) &= r_{i1}(t) - \varepsilon_{i0}\big|r_{i0}(t)-s_i^{\mathrm{DO}}(t)\big|^{\frac12}\operatorname{\sign}\big(r_{i0}(t)-s_i^{\mathrm{DO}}(t)\big), 
% \label{DO:diff1} \\
% \dot r_{i1}(t) &= -\varepsilon_{i1}\operatorname{\sign}\big(r_{i1}(t)-r_{i0}(t)\big),\qquad
% q_i(t):=r_{i0}(t).
% \label{DO:diff2}
% \end{align}
% % Explanation: q_i approximates \dot s_i^{DO}; the differentiator parameters \varepsilon_{i0},\varepsilon_{i1}>0
% % are chosen so that the differentiator residual is small (see Lemma 2 in the SMDO reference). 

% % sliding-mode disturbance observer (scalar/component form)
% \begin{align}
% \dot{\hat{\delta}}_i^{\mathrm{DO}}(t)
% &= -\,q_i(t)
% - a_i\big|\,s_i^{\mathrm{DO}}(t)+q_i(t)\,\big|^{\frac12}\operatorname{\sign}\!\big(s_i^{\mathrm{DO}}(t)+q_i(t)\big) \nonumber \\
% &-b_i\int_{0}^{t}\operatorname{\sign}\!\big(s_i^{\mathrm{DO}(\tau)+q_i(\tau)\big)\,d\tau
% }
% \label{DO:update}
% \end{align}

\subsection{Safety theorem}
 A disturbance observer based RCBF (DORCBF) is proposed in this subsection by explicitly incorporating the disturbane estimate and extending the backstepping technique aiming to propose a control architecture for robust safety-critical control to obtain the persistent safety objective for (\ref{eq:1_f}) with mismatched and matched disturbances defined in (\ref{eq:1_g}).
 \begin{definition}[DORCBF for perturbed system \eqref{eq:1_f}]\label{def:3}
 For the system (\ref{eq:1_f}), a function $h_1(\boldsymbol{\varphi}):\mathbb R^n\to\mathbb R$ belongs to $\mathcal C^{n}$ is a DORCBF on the set \eqref{eq:2_d},
  such that $\forall \boldsymbol{\varphi} \in \mathcal{\hat{\mathcal{S}}}$ and $\forall t \geq 0$ :
\begin{equation}\label{eq:5_d} 
   \sup_{u \in U} \big[\underbrace{{L_{f}h_1(\boldsymbol{\varphi})} + L_{g}h_1(\boldsymbol{\varphi})u - \|h_{\boldsymbol{\varphi}}(\boldsymbol{\varphi})\| \hat{w}}_{\dot{h}_1(\boldsymbol{\varphi},u,\hat{w})} \big]  \geq - \Gamma(h_1({\boldsymbol{\varphi}})),   
\end{equation}
where $\hat{w}=\big[\hat w_1^\top\,\hat w_2^\top\,\cdots\,\hat w_n^\top\big]^\top$ is the estimation of $w=\big[w_1^\top\, w_2^\top\,\cdots\,w_n^\top\,\big]^\top$ in (\ref{eq:1_g}) and  $h_{\boldsymbol{\varphi}}(\boldsymbol{\varphi}) = \frac{\partial h(\boldsymbol{\varphi})}{\partial \boldsymbol{\varphi}}$, we define the point-wise set of controllers:
\begin{equation} \label{eq:dorcbf_set}
    K_{\text{DORCBF}}(\boldsymbol{\varphi}) \triangleq \left\{u \in \mathcal{U} \mid \dot{h}_1(\boldsymbol{\varphi},u,\hat{w}) \geq - \Gamma(h_1(\boldsymbol{\varphi})) \right\}.
\end{equation}
\end{definition}
The requirement for smooth control laws to facilitate the repeated differentiations needed at each stage is a major obstacle when integrating CBF-based techniques with backstepping. These differentiations are incompatible with traditional robust CBF constraints, such as \eqref{eq:5_d}, which are nonsmooth, particularly for higher RD systems~\citep{taylor2022safe}. To resolve this, we present the revised definition of DORCBF by observing the upper bound of the nonsmooth component with a smooth function for $\mu>0$: 
\begin{align}\label{eq:5_aa}   
 \frac{1}{4\mu}\,\|h_{\boldsymbol{\varphi}}(\boldsymbol{\varphi})\|^2 + \mu{\hat{w}}^2 \geq \|h_{\boldsymbol{\varphi}}(\boldsymbol{\varphi})\| {\hat{w}}.
\end{align}
Now introducing a new definition of DORCBFs for \eqref{eq:1_f} as follows. 
\begin{definition}\label{def:3a}
   For the transformed system (\ref{eq:1_f}), a function $h_1(\boldsymbol{\varphi}):\mathbb R^n\to\mathbb R$ belongs to $\mathcal C^{n}$ is a smooth DORCBF (SDORCBF) on the set \eqref{eq:2_d} such that $\forall \boldsymbol{\varphi} \in \hat{\mathcal{S}} $, for $\mu>0$ and $\forall t \geq 0$:
\begin{equation}\label{eq:5_bb} 
   \sup_{u \in U} \Big[\underbrace{ \dot{h}_1(\boldsymbol{\varphi},u) -  \Big( \frac{1}{4\mu}\,\|h_{\boldsymbol{\varphi}}(\boldsymbol{\varphi})\|^2 + \mu{\hat{w}}^2 \Big) }_{\dot{h}_1(\boldsymbol{\varphi},u,\hat{w})} \Big]  \geq - \Gamma(h_1(\mathbf{\varphi})),   
\end{equation}  
we can define the point-wise set of controllers:
\begin{equation} \label{eq:srcbf_set_d}
    K_{\text{SDORCBF}}(\boldsymbol{\varphi}) \triangleq \left\{u \in \mathcal{U} \mid \dot{h}_1(\boldsymbol{\varphi},u,\hat{w}) \geq - \Gamma(h_1(\mathbf{\varphi})) \right\}.
\end{equation}
\end{definition} 
 Now consider a system \eqref{eq:1_f}, and assume a desired CBF $h_1(\boldsymbol{\varphi})$ with RD $n>1$ that also satisfies Assumption \ref{eq:ass_4}. We design a smooth controller that accounts for the disturbances in \eqref{eq:1_g} while evading the singularities linked to nonsmooth constraints by employing the SDORCBF formulation. The time derivative of $h_1(\boldsymbol{\varphi})$ is
% We start by calculating 
\begin{small}
\begin{align}\label{eq:14}
    \dot{h}_1(\boldsymbol{\varphi}) &= L_f h_1(\boldsymbol{\varphi})+\frac{\partial h_1}{\partial \boldsymbol{\varphi}}\hat{w}_1, \nonumber\\
                 &\geq -\varrho_1 \Upsilon(t) h_1(\boldsymbol{\varphi}) + \underbrace{\varrho_1 \Upsilon(t) h_1(\boldsymbol{\varphi}) + L_f h_1(\boldsymbol{\varphi}) - \Lambda_{1}(\boldsymbol{\varphi})}_{h_2(\boldsymbol{\varphi})}, 
                 % \nonumber\\
                 % &\geq -\varrho \Upsilon(t) h_1(x) + \underbrace{\varrho \Upsilon(t) h_1(x) + L_f h_1(x)-\bar{D}}_{h_2(x)},
\end{align}    
\end{small}
where $\Lambda_{1}(\boldsymbol{\varphi})= \frac{1}{4\mu_1}\,\|\frac{\partial h_1}{\partial \boldsymbol{\varphi}}\|^2 + \mu_{1}\hat{w}_1^2$, $\mu_1>0$ and $ \varrho_1 > \max\!\left\{ 0,\; \frac{-L_f h_1(\boldsymbol{\varphi}_0)+\Lambda_{1}(\boldsymbol{\varphi}_0)}{\Upsilon(t_0)\,h_1(\boldsymbol{\varphi}_0)} \right\}$. Then, using the same method as previously mentioned in \eqref{eq:integrator_wo_d} with
the inclusion of backstepping, we determined a CBF as follows:
\begin{small}
\begin{align}\label{eq:11_b}
    h_1(\boldsymbol{\varphi}) &= h_1(\boldsymbol{\varphi}), \nonumber\\
    h_i(\boldsymbol{\varphi}) &= \varrho_{i-1} \Upsilon(t)^{\vartheta(i-1)} h_{i-1}(\boldsymbol{\varphi}) + L_f h_{i-1}(\boldsymbol{\varphi})-\Lambda_{i-1}(\boldsymbol{\varphi}),
\end{align}
\end{small}
for $i=\{2, \ldots, n\}$ and $\Lambda_{i-1}(\boldsymbol{\varphi})= \frac{1}{4\mu_{i-1}}\,\|\frac{\partial h_{i-1}}{\partial \boldsymbol{\varphi}}\|^2 + \mu_{i-1}\hat{w}_{i-1}^2$, while updating the initial gains to include the perturbation bound:
\begin{equation}\label{eq:10_b}
    \varrho_{i-1} > \max\!\left\{ 0,\; \,\frac{-L_f h_{i-1}(\boldsymbol{\varphi}_0)+\Lambda_{i-1}(\boldsymbol{\varphi}_0)}{\Upsilon(t_0)^{\vartheta(i-1)}\,h_{i-1}(\boldsymbol{\varphi}_0)} \right\}.
\end{equation}
As a result, the QP problem and its closed-form solution is formulated as follows:
\begin{align}\label{eq:13_b}  
\boldsymbol{v} &= \arg\min_{\boldsymbol{v}} \; \|\boldsymbol{v} - \boldsymbol{v}_{\mathrm{no}}\|^2 \nonumber \\ 
\text{s.t. } &{\dot{h}_n(\boldsymbol{\varphi},\boldsymbol{v},\hat{w})} -\Lambda_{n}(\boldsymbol{\varphi}) \geq -\varrho_{n} \, \Upsilon(t)^{\vartheta n}\, h_n(\boldsymbol{\varphi}).
\end{align}
\begin{align}\label{eq:18_c} 
\boldsymbol{v} = \begin{cases} 
\boldsymbol{v}_{no}, & \zeta(\boldsymbol{\varphi}, \boldsymbol{v}_{no},\hat{w}) \geq 0, \\ 
\boldsymbol{v}_{no} - (L_g h_n(\boldsymbol{\varphi}))^\top \frac{\zeta(\boldsymbol{\varphi}, \boldsymbol{v}_{no},\hat{w})}{\|L_g h_n(\boldsymbol{\varphi})\|^2}, & \text{otherwise},
\end{cases}
\end{align}
where
\begin{align}\label{eq:19_d} 
\zeta(\boldsymbol{\varphi}, \boldsymbol{v}_{no},\hat{w}) &:= {\dot{h}_n(\boldsymbol{\varphi},\boldsymbol{v}_{no},\hat{w})} -\Lambda_{n}(\boldsymbol{\varphi}) +\varrho_{n} \, \Upsilon(t)^{\vartheta n}\, h_n(\boldsymbol{\varphi}).
\end{align}
\begin{remark}
    Inspired by~\citep{labbadi2024hyperexponential}, our research focuses on time-varying, continuous, and bounded external perturbations that do not depend on system state, such as wind gusts affecting the drone, while the framework in~\citep{kim2025robust} tackles state-dependent perturbations.   This distinction is important because a hovering drone becomes unsafe if disturbances are assumed to disappear at equilibrium. Our approach addresses this gap by ensuring robustness against external disturbances.  
    % This distinction is critical because a controller designed to stabilize a hovering drone would be vulnerable to wind at that position if the disturbance model assumes it vanishes at equilibrium, potentially resulting in a safety breach.
    % \ml{please cite my CSS paper.}
    \hfill$\blacksquare$
\end{remark}

\begin{theorem}\label{th:2}
   For the disturbed system (\ref{eq:1_f}), the safe set $\hat{\mathcal{S}}$ defined in \eqref{eq:2_d}, and the disturbance observer is presented in \ref{eq:DO}. Assume Assumptions~\ref{ass:DO_W_bounds} and~\ref{eq:ass_4} satisfies, and $h_1(\boldsymbol{\varphi}_0) > 0$. Then, the control law (\ref{eq:18_c}), derived using the backstepping transformation (\ref{eq:11_b}) with the initial gain (\ref{eq:10_b}) and $\varrho_n > 0$, ensures that $h_1(\boldsymbol{\varphi}(t))\geq 0$ $\forall t \in [t_0, \infty)$.  
   
\end{theorem}
\section{Illustrative Example}
In this simulation problem setting, we are considering a vehicle and our objective is to navigate safely while being subjected to disturbances. The vehicle is described by:
\begin{align}\label{eq:vehicel_model}
\dot x &= v \cos\theta, \quad \dot y = v \sin\theta, \quad \dot v = u_1, \quad\dot\theta = u_2,    
\end{align}
where the positional states of the vehicle are represented by $[x\, y]^\top$, the turning rate and acceleration are controlled by $u_1$ and $u_2$, respectively. Whereas, $\theta$  and $v$ denote the heading angle and forward speed, respectively, and we assume $v\neq0$. Assume the vehicle is unaware of the dynamics of the obstacle. Because of this, the vehicle will handle the obstacle dynamics as unknown external disturbances, which are illustrated as follows:
\begin{align}\label{eq:obs}
\dot x_d &= d_x, \quad \dot y_d = d_y,
\end{align}
with the position of the obstacle denoted by $(x_d\, y_d)^\top$, and $\boldsymbol{d}=[d_x\, d_y]^\top$
indicates the external disturbances with $\|\boldsymbol{d}\| \le \rho$, $\rho > 0$. By augmenting the vehicle and obstacle dynamics the overall system is as follows:
\[
\dot{\boldsymbol{x}}= \begin{bmatrix} \dot{x} \\ \dot{y} \\ \dot{v} \\ \dot{\theta} \\ \dot{x}_d \\ \dot{y}_d \end{bmatrix}= \begin{bmatrix} v\cos\theta \\ v\sin\theta \\ 0 \\ 0 \\ 0 \\ 0 \end{bmatrix}+\begin{bmatrix}
0 \\
0 \\
u_1 \\
u_2 \\
0 \\
0 
\end{bmatrix} + \begin{bmatrix} 0\\0\\0\\0\\ d_x\\ d_y \end{bmatrix}.
\]
The above dynamics are not in the form of (\ref{eq:1}), to transformed into such a form we define 
\begin{equation*}
\boldsymbol{x}_1 := [x,y]^\top, \qquad \boldsymbol{x}_2 := [v\cos\theta,v\sin\theta]^\top    
\end{equation*}
Then
\begin{align}
\dot{\boldsymbol{x}}_1 &= \boldsymbol{x}_2 + \underbrace{\phi_1(\bar{\boldsymbol{x}}_1)}_{0}, \nonumber\\
    \dot{\boldsymbol{x}}_2 &=G(x)u +\underbrace{\phi_2(\bar{\boldsymbol{x}}_2)}_{0},
\end{align}
where 
\begin{align}
    G(x)&= \begin{bmatrix}
\cos\theta & -v\sin\theta\\
\sin\theta & v\cos\theta
\end{bmatrix}, \qquad  u= \begin{bmatrix}
u_1\\
u_2.
\end{bmatrix}
\end{align}
Now, applying the coordinate transformation gives ${\boldsymbol{\beta}}_0=0, {\boldsymbol{\beta}}_1 = -\phi_1 = 0$ and ${\boldsymbol{\beta}}_2= -\phi_2 + \frac{\partial\beta_1}{\partial x_1}(x_2+\phi_1)=0$. The transformed coordinates are simply
\begin{align}
 {\boldsymbol{\varphi}}_1 &= \boldsymbol{x}_1 := \begin{bmatrix} x\\ y\end{bmatrix}=\begin{bmatrix} {\varphi}_{1,1}\\ {\varphi}_{1,2} \end{bmatrix},  \nonumber\\
{\boldsymbol{\varphi}}_2 &= 
\boldsymbol{x}_2 := \begin{bmatrix} v\cos\theta\\ v\sin\theta \end{bmatrix}=\begin{bmatrix} {\varphi}_{2,1}\\ {\varphi}_{2,2} \end{bmatrix},   
\end{align}
The transformed chain becomes the exact vector chain
\begin{equation}\label{eq:chain}
\dot{\boldsymbol{\varphi}}_1 = {\boldsymbol{\varphi}}_{2} , \quad
\dot{\boldsymbol{\varphi}}_2 = \boldsymbol{v}
\end{equation}
where ${\boldsymbol{v}}= ({v}_{1},{v}_{2})^{T}\in\mathbb{R}^2$ is the new transformed input. The original actuator inputs are obtained by $u = G^{-1}(x){\boldsymbol{v}}$. To incorporate the moving obstacle, the dynamics were modeled as another vehicle:
\begin{align}\label{eq:obs_model}
\dot x_d &= v_d \cos\theta, \quad \dot y_d = v_d \sin\theta, \quad \dot\theta_d = u_d,    
\end{align}
define the relative position:
\begin{equation}\label{eq:ex2_delta_def}
\boldsymbol{\Delta} := {\boldsymbol{\varphi}}_{1} - \boldsymbol{p}_d,\qquad \boldsymbol{p}_d=[ x_d, \; y_d]^\top,
\end{equation}
where $\boldsymbol{p}_d$ denotes the obstacle vehicle position. Differentiating \eqref{eq:ex2_delta_def} and using \eqref{eq:chain} yields
\begin{equation}\label{eq:ex2_delta_dot}
\dot{\boldsymbol{\Delta}}=\dot{\boldsymbol{\varphi}}_1 -\dot {\boldsymbol{p}}_d={\boldsymbol{\varphi}}_{2}-\boldsymbol{d},\qquad
\boldsymbol{d}=\dot {\boldsymbol{p}}_d=[\dot x_d,\; \dot y_d]^\top.
\end{equation}
Treating the velocity of the obstacle vehicle as a disturbance $w_1 := -\boldsymbol{d}$, we obtain the perturbed chain form
\begin{equation}\label{eq:ex2_chain_dist}
\dot{\boldsymbol{\varphi}}_1 = {\boldsymbol{\varphi}}_{2}+w_1  , \quad
\dot{\boldsymbol{\varphi}}_2 = \boldsymbol{v}+w_2,
\end{equation}
where for the moving obstacle vehicle only scenario $w_2=0$. To circumvent obstacle collision, a candidate CBF is considered as:
\begin{align}\label{eq:cbf_obs}
 h_1({\boldsymbol{\varphi}}_1) &= ({\varphi}_{1,1}-x_d)^2 + ({\varphi}_{1,2}-y_d)^2 - r^2,   
\end{align}
while, $r \in \mathbb{R}$ defines the user-specified safety distance. Differentiating \eqref{eq:cbf_obs} and employing the disturbance estimate $\hat w_1$ from the observer yields 
% Differentiate to form the backstepping second-level CBF \(h_2\). Let \(\Delta_1 := {\varphi}_{1,1}-x_d,\ \Delta_2 := {\varphi}_{1,2}-y_d\), $\boldsymbol{\Delta}=[\Delta_1,\Delta_2]^\top$ and $\boldsymbol{d}=[d_x,d_y]^\top$. Also $\dot{\boldsymbol{\Delta}} = \boldsymbol{\varphi}_2 - \boldsymbol{d}$.
\begin{align}
 \dot h_1 ({\boldsymbol{\varphi}}_1) &= 2({\varphi}_{1,1}-x_d)\,({\varphi}_{2,1}) + 2({\varphi}_{1,2}-y_d)\,({\varphi}_{2,2}) \nonumber\\
 &\quad -2({\varphi}_{1,1}-x_d)\,d_x - 2({\varphi}_{1,2}-y_d)\,d_y \nonumber\\
 &= 2\boldsymbol{\Delta}^\top\boldsymbol{\varphi}_2 + 2\boldsymbol{\Delta}^\top\hat w_1 \nonumber\\
 &\geq -\varrho_1 \Upsilon(t) h_1(\boldsymbol{\varphi}_1)\nonumber\\
 & \quad + \underbrace{\varrho_1 \Upsilon(t) h_1(\boldsymbol{\varphi}_1) + L_f h_1(\boldsymbol{\varphi}_1) - \Lambda_{1}(\boldsymbol{\varphi}_1)}_{h_2(\boldsymbol{\varphi})},
\end{align}
and
\begin{align}
  h_2(\boldsymbol{\varphi})&={\varrho_1 \Upsilon(t) h_1(\boldsymbol{\varphi}_1) + L_f h_1(\boldsymbol{\varphi}_1) - \Lambda_{1}(\boldsymbol{\varphi}_1)} , 
\end{align}
whereas, $L_f h_1(\boldsymbol{\varphi}_1)=2\boldsymbol{\Delta}^\top\boldsymbol{\varphi}_2$ and $\Lambda_{1}(\boldsymbol{\varphi}_1)= \frac{1}{4\mu_1}\,\|\frac{\partial h_1}{\partial \boldsymbol{\varphi}_1}\|^2 + \mu_{1}\|\hat w_1\|^2=\frac{1}{\mu_1}\|\Delta\|^2
+\mu_1\|\hat w_1\|^2$, $\mu_1>0$. Now, we select $\varrho_1 > \max\!\left\{ 0,\; \frac{-L_f h_1(\boldsymbol{\varphi}_0)+\Lambda_{1}(\boldsymbol{\varphi}_0)}{\Upsilon(t_0)\,h_1(\boldsymbol{\varphi}_0)} \right\}$, such that $h_2(\boldsymbol{\varphi}_0)>0$. Theorem \ref{th:2} demonstrates that imposing $h_2(\boldsymbol{\varphi})>0$ guarantees the safety of the system. Computing the time derivative of $h_2(\boldsymbol{\varphi})>0$ results in
\begin{small}
\begin{align}\label{eq:h2_dot}
\dot h_2=&\;\varrho_1\vartheta\Upsilon^{\vartheta-1}h_1
+2\varrho_1\,\Upsilon(t)^{\vartheta}\,\boldsymbol{\Delta}^\top(\boldsymbol{\varphi}_2+\hat w_1)
+2\|\boldsymbol{\varphi}_2+\hat w_1\|^2 \nonumber\\
&\;+2\boldsymbol{\Delta}^\top \boldsymbol{v}
+2\boldsymbol{\Delta}^\top\dot{\hat w}_1
-\frac{2}{\mu_1}\boldsymbol{\Delta}^\top(\boldsymbol{\varphi}_2+\hat w_1)
-2\mu_1\hat w_1^\top\dot{\hat w}_1.
\end{align}    
\end{small}
\begin{align}
\boldsymbol{v}
&= \arg\min_{\boldsymbol{v}} \;\|\boldsymbol{v}-\boldsymbol{v}_{\mathrm{no}}\|^2 \nonumber\\
\text{s.t.}\qquad & \dot h_2(\boldsymbol{\varphi},\boldsymbol{v}_{no},\hat w_1,\dot{\hat w}_1) \ge -\varrho_2\,\Upsilon^{2\vartheta}h_2.
\end{align}
 The proposed safety controller is applied on the vehicle and named as DORCBF compared while on another vehicle standard backstepping CBF referred to as BCBF strategy outlined in \citep{krstic2006nonovershooting} is applied to test the safety performance subjected to unknown external perturbations. The safe distance is selected as $r=1$ and both vehicles are initialized at $[x(0) \,y(0) \, v(0)  \, \theta(0)]^\top=[0 \,0 \, 0  \, 0]^\top$, while the obstacle vehicle is initialized at $[x_d(0) \, y_d(0) \, \theta_d(0)]^\top=[3 \,-3 \, \frac{\pi}{2}]^\top$, with $v_d(t)=1$ and $u_d(t)=2\cos(2t)$. The nominal control aim of the vehicle was to drive to the right at a constant velocity of $v=1$ with a heading angle $\theta=0$. The nominal control commands were therefore specified as $u_{1(no)} = -a_1(v-1)$ and $u_{2(no)}=-a_2 \theta$, where $a_1 = a_2 = 1$. The parameters were chosen as $\varrho_1=5$, $\varrho_2=0.5$, $\mu_1=0.2$, and $\vartheta=3$. The gains in DO design were selected as $\lambda_{10}=20$, $\lambda_{11}=10$, and $k_{11}=k_{12}=10$. 

\begin{figure}[]
    \centering
    \includegraphics[scale=0.45]{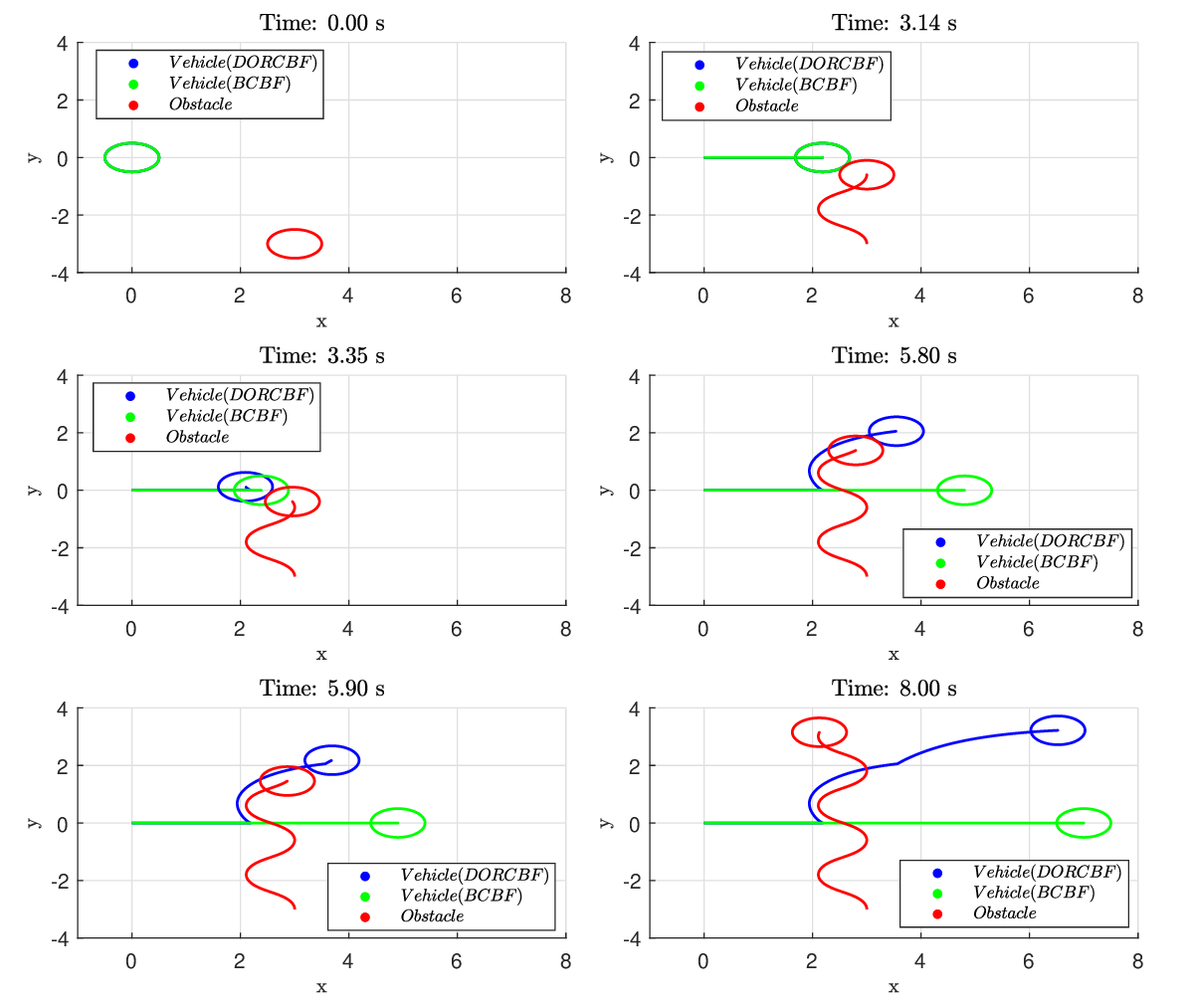}
    \caption{System trajectory of vehicles in the presence of an
obstacle (red) with unknown dynamics. The vehicle using
the standard BCBF (green) fails to evad collision, while the
vehicle considering the proposed safety filter DORCBF (blue) remains
safe from the obstacle by effectively guiding the vehiclem.}
    \label{fig:result1}
\end{figure}

\begin{figure}[]
    \centering
    \includegraphics[scale=0.58]{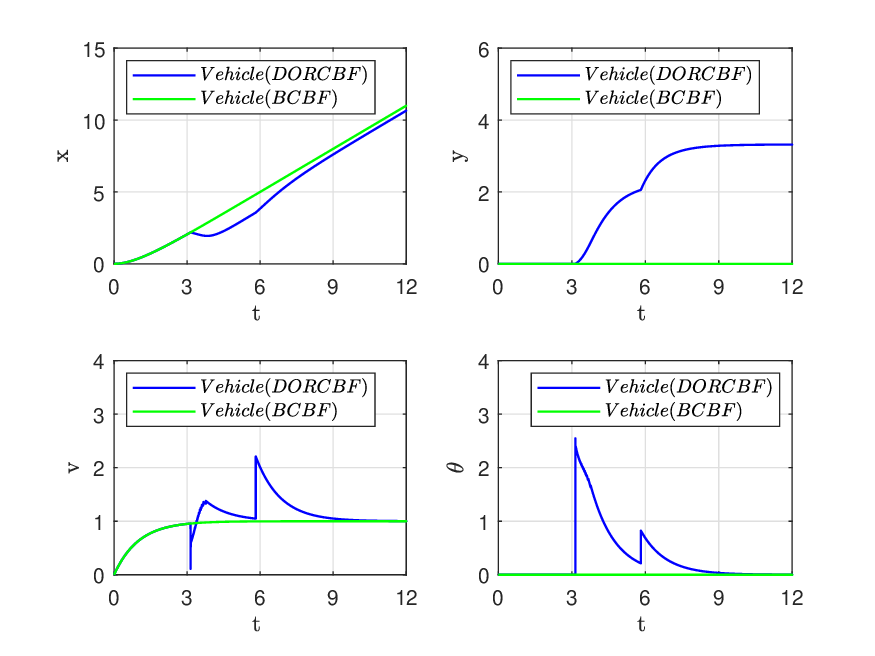}
    \caption{ System states of the vehicles representing the positional states $[x\,  y]$, forward velocity $v$, and the heading
angle $\theta$.}
    \label{fig:result2}
\end{figure}

The corresponding system trajectories are shown in Fig. \ref{fig:result1}, and the evolution of the vehicle states under the standard BCBF strategy and the DORCBF framework is shown in Fig. \ref{fig:result2}. As can be observed at $t = 3.35$ sec in Fig. \ref{fig:result1}, the vehicle using the BCBF technique breaches safety by failing to account for the obstacle's dynamics, which leads to a collision. On the other hand, the vehicle that incorporates external perturbations and employs the DORCBF prevents collisions and keeps a safe distance from dynamic obstacle when it encounters them twice, roughly at $t = 3.35$ sec and $t = 5.80$ sec, validating our method.

% {\color{blue}
% \subsection{\bf{Example 1 taken from} \citep{abel2023prescribed}} 
% To illustrate our proposed design on a vector-valued chain of integrators with disturbances, we consider the following system:
% \begin{align}\label{eq:1_f_b}
% \dot{\boldsymbol{\varphi}}_1 &= {\boldsymbol{\varphi}}_{2} + \boldsymbol{d}_1, \nonumber\\
% \dot{\boldsymbol{\varphi}}_2 &= \boldsymbol{v} + \boldsymbol{d}_2,
% \end{align}
% where ${\boldsymbol{\varphi}}_1= ({\varphi}_{1,1},{\varphi}_{1,2})^{T}$, ${\boldsymbol{\varphi}}_2= ({\varphi}_{2,1},{\varphi}_{2,2})^{T}$, are vector-valued
% states in $\mathbb{R}^2$, while ${\boldsymbol{d}}_1= ({d}_{1,1},{d}_{1,2})^{T}$, ${\boldsymbol{d}}_2= ({d}_{2,1},{d}_{2,2})^{T}$, are vector-valued
% mismatched and matched disturbances in $\mathbb{R}^2$ are introduced, and ${\boldsymbol{v}}= ({v}_{1},{v}_{2})^{T}$ is the vector-valued input. A
% nominal controller is designed to stabilize the states $({\boldsymbol{\varphi}}_1,{\boldsymbol{\varphi}}_2)$
% to the origin, while the safety condition is to keep the state ${\boldsymbol{\varphi}}_1$ outside of a circular unsafe set, that is, 
% \begin{align}\label{eq:safe_set}
% h({\boldsymbol{\varphi}}_1)&=({\varphi}_{1,1}-c_{1})^2 + ({\varphi}_{1,2}-c_{2})^2 -r^{2}
% \end{align}
% where $(c_1,c_2) \in \mathbb{R}^2$ and $r \in \mathbb{R}$ are the center and radius of the
% unsafe-set, respectively. Now apply the proposed safety filter on this example and compare the results with other techniques.

\section{Conclusions}
In this paper, a robust approach for ensuring the safety of strict-feedback systems is proposed. The proposed method is applied both in the absence of disturbances and in the presence of matched and mismatched disturbances. A linear time-varying control law is designed using a backstepping approach within the CBF framework. The theoretical guarantees are established based on the convergence of the designed observer, time-scale transformations, and quadratic programming (QP) methods. An illustrative example is provided to demonstrate the effectiveness of the proposed approach.
\appendix{}
\section{Proofs}
\subsection{Proof of Proposition~\ref{pro:2}}
\begin{proof}
   Given $h_{i-1}(\boldsymbol{\varphi}_0) > 0$, the initial gain in \eqref{eq:10_a} is intended to ensure $h_i(\boldsymbol{\varphi}_0) > 0$. Since we know $h_1(\boldsymbol{\varphi}_0) > 0$, choosing 
\begin{equation}\label{eq:10_aa}
   \bar{\varrho}_1 > \max\!\left\{ 0,\; -\,\frac{L_f h_1(\boldsymbol{\varphi}_0)}{\Upsilon(t_0)\,h_1(\boldsymbol{\varphi}_0)} \right\},
\end{equation}
guarantees that $h_2(\boldsymbol{\varphi}_0) > 0$. Thus, reiterating this argument inductively yields $h_i(\boldsymbol{\varphi}_0) > 0$ for $i=\{2, \ldots, n\}$. Then, by differentiating the CBF recursion in \eqref{eq:11_a} yield: 
\begin{align}\label{eq:28} 
\frac{d}{dt}h_k(\boldsymbol{\varphi}(t))&=-\bar{\varrho}_{k} \Upsilon(t)^{\vartheta k}h_{k}(\boldsymbol{\varphi}(t)) + h_{k+1}(\boldsymbol{\varphi}(t)),
\end{align}
\begin{align}\label{eq:29} 
\frac{d}{dt}h_n(\boldsymbol{\varphi}(t))&=L_f h_n(\boldsymbol{\varphi}(t))+L_g h_n(\boldsymbol{\varphi}(t))u \nonumber\\
                    &\geq  -\bar{\varrho}_{n} \, \Upsilon(t)^{\vartheta n}\, h_n(\boldsymbol{\varphi}(t)),
\end{align}
for $k=\{1, \ldots, n-1\}$.
Employing the  comparison lemma and the variation of constants formula, for $t \in [t_0, \infty)$, gives

\begin{align}\label{eq:30}
  h_{k}(\boldsymbol{\varphi}(t)) &\ge h_{k}(\boldsymbol{\varphi}({t_0})) e^{-\bar{\varrho}_k \int_{t_{0}}^{t} \Upsilon(s)^{\vartheta k}\,ds} \nonumber\\
&\quad + \int_{t_{0}}^{t} h_{k+1}(s) e^{-\bar{\varrho}_k \int_{t_0}^{t} \Upsilon(\tau)^{\vartheta k}\,d\tau} ds,  
\end{align}

\begin{small}
\begin{align}\label{eq:31}
  h_{n}(\boldsymbol{\varphi}(t)) &\ge h_{n}(\boldsymbol{\varphi}(t_{0})) e^{-\bar{\varrho}_n \int_{t_{0}}^{t} \Upsilon(s)^{\vartheta n}\,ds}, 
\end{align}    
\end{small}
for $k=\{1, \ldots, n-1\}$. Because, $h_n(\boldsymbol{\varphi}(t_0))>0$, inequality \eqref{eq:31} ensures $h_n(\boldsymbol{\varphi}(t))>0$ for all $t \in [t_0, \infty)$. Substituting \eqref{eq:31} into \eqref{eq:30} for $k=n-1$ leads to:
\begin{align}\label{eq:32}
  h_{n-1}(\boldsymbol{\varphi}(t)) &\ge \underbrace{h_{n-1}(\boldsymbol{\varphi}({t_0})) e^{-\bar{\varrho}_{n-1} A(t)}}_{>0} \nonumber\\
& + \underbrace{ h_n(\boldsymbol{\varphi}(t_0)) e^{-\bar{\varrho}_{n-1} A(t)} \int_{t_0}^{t} e^{\bar{\varrho}_{n-1} A(\tau) - \varrho_{n} B(\tau) } d\tau}_{ \geq 0 }, \nonumber \\
& \geq h_{n-1}(\boldsymbol{\varphi}({t_0})) e^{-\bar{\varrho}_{n-1} A(t)} > 0,
\end{align}
where $\bar{A}(t)=\int_{t_0}^{t} \Upsilon(s)^{\vartheta(n-1)}  ds$ and $\bar{B}(t)=\int_{t_0}^{t} \Upsilon(s)^{\vartheta n}  ds$. Employing backward induction with (\ref{eq:31}) and (\ref{eq:32}),
we finally obtain
\begin{align}\label{eq:33}
  h_{1}(\boldsymbol{\varphi}(t)) &\ge h_{1}(\boldsymbol{\varphi}(t_{0})) e^{-\bar{\varrho}_1 \int_{t_{0}}^{t} \Upsilon(s)\,ds}, 
\end{align} 
which implies
\begin{align}\label{eq:34}
  h_{1}(\boldsymbol{\varphi}(t)) &\ge  0, \quad \forall t \in [t_0, \infty). 
\end{align} 
\end{proof}
\subsection{Proof of Proposition~\ref{prop:DO_vector}}
\begin{proof}
For $i\in\{1,\dots,n\}$, the auxiliary variable in the disturbance estimator
and the estimation error from \eqref{eq:sigma10_dot}. The differentiator approximation and bounded residual property is given by \eqref{eq:fosmd}.  
Consequently, \eqref{eq:fosmd_3} provides the dynamics of the sliding variable $\sigma_{i1}$, with $\bar D_i$ specified in \eqref{eq:fosmd_3}. By Assumption~\ref{ass:DO_W_bounds} and the differentiator approximation, the bound \eqref{eq:Di_bar_bound} holds. Hence, Lemma~\ref{eq:lemma_1} applies to the dynamics
in \eqref{eq:fosmd_3}. Under the gain conditions \eqref{eq:k_gains_prop}, it follows that $\sigma_{i1} \rightarrow 0$ (and $\dot{\sigma}_{i1}\rightarrow 0$) in finite time. Thus, $\sigma_{i1} = \sigma_{i0} + \zeta_i \implies$ the signal $\dot{\sigma}_{io}$ is convergent in finite time according
to \eqref{eq:fosmd}. Using \eqref{eq:sigma10_dot}, this implies $\tilde w_i\to 0$ in finite time. 
\end{proof}
\subsection{Proof of Theorem~\ref{th:2}}
\begin{proof}
    Given $h_{i-1}(\boldsymbol{\varphi}_0) > 0$, the initial gain (\ref{eq:10_b}) is selected to ensure $h_i(\boldsymbol{\varphi}_0) > 0$. Because $h_1(\boldsymbol{\varphi}_0) > 0$, choosing
\begin{align}\label{eq:35}
     \varrho_1 > \max\!\left\{ 0,\; \frac{-L_f h_1(\boldsymbol{\varphi}_0)+\Lambda_{1}(\boldsymbol{\varphi}_0)}{\Upsilon(t_0)\,h_1(\boldsymbol{\varphi}_0)} \right\},
\end{align}
guarantees $h_2(\boldsymbol{\varphi}_0) > 0$. Repeating this argument for $i=\{2, \ldots, n\}$, it follows that $h_i(\boldsymbol{\varphi}_0) > 0$ by induction. Differentiating the recursive CBFs in \eqref{eq:11_b} yields: 
\begin{align}\label{eq:28_a} 
\frac{d}{dt}h_k(\boldsymbol{\varphi}(t))&=-\varrho_{k} \Upsilon(t)^{\vartheta k}h_{k}(\boldsymbol{\varphi}(t)) + h_{k+1}(\boldsymbol{\varphi}(t)),
\end{align}
\begin{align}\label{eq:29_a} 
\frac{d}{dt}h_n(\boldsymbol{\varphi}(t))&=L_f h_n(\boldsymbol{\varphi}(t))+L_g h_n(\boldsymbol{\varphi}(t))u-\Lambda_{n}(\boldsymbol{\varphi}) \nonumber\\
                    &\geq  -\varrho_{n} \, \Upsilon(t)^{\vartheta n}\, h_n(\boldsymbol{\varphi}(t)),
\end{align}
for $k=\{1, \ldots, n-1\}$. Using the variation of constants formula and the comparison lemma, for $t \in [t_0, \infty)$, gives
\begin{align}\label{eq:30_a}
  h_{k}(\boldsymbol{\varphi}(t)) &\ge h_{k}(\boldsymbol{\varphi}({t_0})) e^{-\varrho_k \int_{t_{0}}^{t} \Upsilon(s)^{\vartheta k}\,ds} \nonumber\\
&\quad + \int_{t_{0}}^{t} h_{k+1}(s) e^{-\varrho_k \int_{t_0}^{t} \Upsilon(\tau)^{\vartheta k}\,d\tau} ds,  
\end{align}
\begin{small}
\begin{align}\label{eq:31_a}
  h_{n}(\boldsymbol{\varphi}(t)) &\ge h_{n}(\boldsymbol{\varphi}(t_{0})) e^{-\varrho_n \int_{t_{0}}^{t} \Upsilon(s)^{\vartheta n}\,ds}, 
\end{align}    
\end{small}
for $k=\{1, \ldots, n-1\}$. As previously stated, $h_n(\boldsymbol{\varphi}(t_0))>0$. Consequently, inequality \eqref{eq:31_a} ensures $h_n(\boldsymbol{\varphi}(t))>0$ $\forall t \in [t_0, \infty)$. Substituting (\ref{eq:31_a}) into (\ref{eq:30_a}) for $k=n-1$ yields
\begin{align}\label{eq:32_a}
  h_{n-1}(\boldsymbol{\varphi}(t)) &\ge \underbrace{h_{n-1}(\boldsymbol{\varphi}({t_0})) e^{-\varrho_{n-1} A(t)}}_{>0} \nonumber\\
& + \underbrace{ h_n(\boldsymbol{\varphi}(t_0)) e^{-\varrho_{n-1} A(t)} \int_{t_0}^{t} e^{\varrho_{n-1} A(\tau) - \varrho_{n} B(\tau) } d\tau}_{ \geq 0 }, \nonumber \\
& \geq h_{n-1}(\boldsymbol{\varphi}({t_0})) e^{-\varrho_{n-1} A(t)} > 0,
\end{align}
where $A(t)=\int_{t_0}^{t} \Upsilon(s)^{\vartheta(n-1)}  ds$ and $B(t)=\int_{t_0}^{t} \Upsilon(s)^{\vartheta n}  ds$. Employing backward induction with (\ref{eq:31_a}) and (\ref{eq:32_a}),
it follows that
\begin{align}\label{eq:33_a}
  h_{1}(\boldsymbol{\varphi}(t)) &\ge h_{1}(\boldsymbol{\varphi}(t_{0})) e^{-\varrho_1 \int_{t_{0}}^{t} \Upsilon(s)\,ds}, 
\end{align} 
hence
\begin{align}\label{eq:34_a}
  h_{1}(\boldsymbol{\varphi}(t)) &\ge  0, \quad \forall t \in [t_0, \infty). 
\end{align} 
\end{proof}

\bibliography{bib}

\end{document}